\documentclass[12pt]{iopart}

\usepackage{bm}
\usepackage{mathrsfs}
\usepackage{verbatim}
\usepackage[all,cmtip]{xy}
\usepackage{amssymb}
\usepackage{graphics}
\usepackage{amsthm}
\usepackage{color}
\usepackage{amscd}
\usepackage{cases}

\usepackage{amssymb}
\usepackage{color}
\usepackage{dsfont}
\definecolor{darkred}  {rgb}{0.5,0,0}
\definecolor{darkblue} {rgb}{0,0,0.5}
\definecolor{darkgreen}{rgb}{0,0.5,0}
\usepackage{hyperref}
\hypersetup{
	pdftitle = {QRT Proposal},
	pdfauthor = {},
	colorlinks = true,
	urlcolor  = blue,         
	linkcolor = red,     
	citecolor = blue,    
	filecolor = darkred       
}

\def\ra{\rangle}
\def\la{\langle}

\def\ot{\otimes}
\def\ra{\rangle}
\def\la{\langle}

\newtheorem{theorem}{Theorem}

\newtheorem{pro}[theorem]{Proposition}

\theoremstyle{remark}

\newcommand{\beax}{\begin{eqnarray*}}
	\newcommand{\eeax}{\end{eqnarray*}}

\def\be{\begin{eqnarray}}
	\def\ee{\end{eqnarray}}
\newcommand{\bea}{\begin{eqnarray}}
	\newcommand{\eea}{\end{eqnarray}}

\newcommand{\mH}{\mathcal{H}}

\newcommand{\mS}{\mathcal{S}}

\newcommand{\lr}{\rangle\langle}

\newcommand{\C}{\mathbb{C}}

\def\>{\rangle}
\def\<{\langle}

\def\diag{ \mathrm{diag}}

\begin{document}

\title[Partial-Norm of Entanglement]{Partial-Norm of Entanglement: Entanglement Monotones That are not Monogamous}

\author{Yu Guo}

\address{School of Mathematical Sciences, Inner Mongolia University, Hohhot, Inner Mongolia 010021, People's Republic of China}
\ead{guoyu3@aliyun.com}

\begin{abstract}
Quantum entanglement is known to be monogamous, i.e., it obeys
strong constraints on how the entanglement can be distributed among multipartite systems. Almost all the entanglement monotones so far are shown to be monogamous. We explore here a family of entanglement monotones with the reduced functions are concave but not strictly concave and show that they are not monogamous. They are defined by four kinds of the ``partial-norm'' of the reduced state, which we call them \textit{partial-norm of entanglement}, minimal partial-norm of entanglement, reinforced minimal partial-norm of entanglement, and \textit{partial negativity}, respectively.
This indicates that, the previous axiomatic definition of the entanglement monotone
needs supplemental agreement that the reduced function should be strictly concave since such a strict concavity can make sure that the corresponding convex-roof extended entanglement monotone is monogamous.
Here, the reduced function of an entanglement monotone refers to the corresponding function on the reduced state for the measure on bipartite pure states. 

\vspace{2pc}
\end{abstract}


\maketitle

Entanglement, as a quintessential manifestation
of quantum mechanics~\cite{Nielsen,Schrodinger1935,Schrodinger1936},
has shown to be a crucial resource in various quantum information processing tasks~\cite{Nielsen,Bennett1992communication,Bennett1993teleporting,Bennett1996prl,Zhang2006experimental}.
The most striking property of entanglement is its distributability,
that is, the impossibility of
sharing entanglement unconditionally across many subsystems
of a composite quantum system~\cite{Osborne,Terhal2004}. 
Understanding how entanglement can be quantified and distributed over many parties reveals fundamental
insights into the nature of quantum correlations~\cite{Horodecki2009} and has profound
applications in both quantum communication~\cite{Bennett2014,Toner,Seevinck} and other area of physics~\cite{streltsov2012are,Augusiak2014pra,Ma2011,Brandao2013,Garcia,Bennett2014,Lloyd}.
Particularly, monogamy law of quantum correlation is the predominant feature 
that guarantees the quantum key distribution secure~\cite{Terhal2004,Pawlowski}.

Quantitatively, the monogamy of entanglement is described by an inequality,
involving a bipartite entanglement monotone. The term
``monotone'' refers to the fact that a proper measure of
entanglement cannot increase on average under local operations and classical communication (LOCC)~\cite{Vidal2000,Vedral1998pra,Plenio2005prl}.
Recall that the traditional monogamy relation of entanglement 
measure $E$ is quantitatively displayed as an inequality of the following form:
\bea\label{basic}
E\left(A|BC\right)\geq E\left(AB\right)+E \left(AC\right),
\eea
where the vertical bar indicates the bipartite split 
across which  the (bipartite) entanglement is measured. 
However, Eq.~(\ref{basic}) is not valid for many entanglement measures but $E^\alpha$
satisfies the relation for some $\alpha>0$~\cite{Coffman,Osborne,Bai}. 
Intense research has been undertaken in this
direction. It has been proved that the squashed entanglement and the one-way distillable entanglement are monogamous~\cite{Koashi}, and almost all the bipartite entanglement measures so far
are monogamous for the multiqubit system or monogamous on pure states~\cite{Coffman,Osborne,streltsov2012are,
	Bai,Luo2016pra,Dhar,Hehuan}.
However, for the higher dimensional system, it is difficult to check the monogamy of entanglement measure
according to Eq.~(\ref{basic}) in general. Consequently, the definition of the monogamy is then improved as~\cite{GG}:
a measure of entanglement $E$ is monogamous if for any $\rho^{ABC}\in\mS^{ABC}$
that satisfies the \textit{disentangling condition}, i.e.,
\bea\label{cond}
E\left( A|BC\right) =E\left( AB\right),
\eea
we have that $E(AC)=0$, where $\mS^X$ denotes the set of all density matrices acting on the state space $\mH^X$. It is equivalent to the traditional monogamy relation in Eq.~(\ref{basic}) for any continuous measure $E$~\cite{GG}: a continuous measure $E$ is monogamous according to this definition if and only if
there exists $0<\alpha<\infty$ such that
$E^\alpha\left( A|BC\right) \geq E^\alpha\left( AB\right) +E^\alpha\left( AC\right)$,	
for all $\rho^{ABC}\in\mS^{ABC}$ 
with fixed $\dim\mH^{ABC}=d<\infty$. 
Such a definition simplifies the justification of the monogamy of entanglement measure greatly~\cite{GG,GG2019}.

Recall that, a function $E: \mS^{AB}\to\mathbb{R}_{+}$ is called a 
measure of entanglement if {(1)} $E(\sigma^{AB})=0$ for any 
separable density matrix $\sigma^{AB}\in\mS^{AB}$, and {(2)} 
$E$ behaves monotonically under LOCC. 
Moreover, convex measures of entanglement that do not increase \emph{on average}
under LOCC are called entanglement monotones~\cite{Vidal2000}.
Let $E$ be a measure of entanglement on bipartite states.  
We define
$E_F\left(\rho^{AB}\right)\equiv\min\sum_{j=1}^{n}p_jE\left(|\psi_j\lr\psi_j|^{AB}\right)$,
where the minimum is taken over all pure state decompositions of $\rho^{AB}=\sum_{j=1}^{n}p_j|\psi_j\lr\psi_j|^{AB}$.
That is, $E_F$ is the convex roof extension of $E$.
Vidal~\cite[Theorem 2]{Vidal2000} showed that for any entanglement measure $E$, $E_F$ above is an entanglement monotone 
if 
\bea\label{h}
h\left( \rho^A\right) = E\left( |\psi\lr\psi|^{AB}\right)
\eea 
is concave, i.e.
$h[\lambda\rho_1+(1-\lambda)\rho_2]\geq\lambda h(\rho_1)+(1-\lambda)h(\rho_2)$
for any states $\rho_1$, $\rho_2$, and any
$0\leq\lambda\leq1$. 
Hereafter, we call $h$ the \textit{reduced function} of $E$ and $\mH^A$ the \textit{reduced subsystem} for convenience.

In Ref.~\cite{GG}, according to definition~(\ref{cond}), we showed that $E_F$ is monogamous whenever $E_F$ is defined via Eq.~(\ref{h}) with $h$ is strictly concave additionally. Except for the R\'enyi $\alpha$-entropy of entanglement with $\alpha>1$, all other measures of entanglement, that were studied intensively in literature, correspond on pure bipartite state to strict concave functions of the reduced density matrix. Theses include the original entanglement of formation~\cite{Bennett1996pra}, tangle~\cite{Rungta2003pra}, concurrence\cite{Rungta,Rungta2003pra}, $G$-concurrence~\cite{Gour2005pra}, Tsallis entropy of entanglement~\cite{Kim2010pra}, and the entanglement measures induced by the fidelity distances~\cite{Guo2020qip}. 
Nevertheless, 
we are not sure yet whether the entanglement monotone is monogamous if the reduced function
is concave but not strictly concave.
The purpose of this paper is to address such a issue. 
We explore the entanglement monotone suggested in Ref.~\cite{Vidal1999pral}, from which we also obtain another two
entanglement monotones. We also investigate the partial negativity which is defined as the norm of the negative part of the state after partial transposition. The reduced functions of theses quantities are not strictly concave, and they are not equivalent to each other. We then show that they are not monogamous. This is the first time to prove that there exist entanglement monotones that are not monogamous in the light of the disentangling condition. 
Our results establish a more closer relation between the monogamy of an entanglement monotone and the strict concavity of the reduced function and suggest that we should require the strict concavity of the reduced function for any ``fine'' entanglement monotone. Moreover, comparing with other reduced functions for which the corresponding entanglement measures are shown to be monogamous, we find that if the reduced function is defined on all of the eigenvalues of the reduced state it is strictly concave and vice versa in general.

Let $|\psi\ra=\sum_{j=1}^r\lambda_j|e_j\ra^A|e_j\ra^B$
be the Schmidt decomposition of $|\psi\ra\in\mH^{AB}$, where $\lambda_1\geq\lambda_2\geq\cdots\geq\lambda_r$, and $r$ is the Schmidt rank of $|\psi\ra$.
In 1999, Vidal proposed an entanglement monotone in Ref.~\cite{Vidal1999pral}, i.e.,
\bea
E_k\left( |\psi\ra\right) =\sum\limits_{i=k}^r\lambda_i^2,\quad k\geq2.
\eea 
In particular,
\bea
E_2\left( |\psi\ra\right) =\sum\limits_{i=2}^r\lambda_i^2=1-\lambda_1^2=1-\|\rho^A\|,
\eea 
where $\rho^A=\tr_B|\psi\ra\la\psi|$, $\|\cdot\|$ is the operator norm, i.e.,
$\|X\|=\sup_{|\psi\ra}\|X|\psi\ra\|$.
Hereafter, we call $E_2$ the \textit{partial-norm of entanglement} in the sense that $1-\|\rho^A\|$ counts for only a portion of the norm $\|\rho^A\|$ for the qubit case.
Obviously, $E_2\geq0$ for any $|\psi\ra\in\mH^{AB}$ and
$E_2(|\psi\ra)=0$ if and only if $|\psi\ra$ is separable.
For mixed state, $E_2(\rho)$ is defined by the convex-roof extension.
Generally, 
$\|A+B\|=\|A\|+\|B\|$ does not guarantee $A=\alpha B$ for hermitian operators $A$ and $B$,
so this reduced function $h(\rho)=1-\|\rho\|$ is not strictly concave.
We next illustrate with counter-examples that $E_2$ is not monogamous.

\begin{theorem}
	$E_2$ is not monogamous.
\end{theorem}

Let 
\beax
|\psi_0\ra^{AB}&=&a_0|0\ra^A|0\ra^B+a_1|1\ra^A|1\ra^B+a_2|2\ra^A|2\ra^B,\\
|\psi_1\ra^{AB}&=&a'_0|0\ra^A|3\ra^B+a'_1|1\ra^A|2\ra^B+a'_2|2\ra^A|1\ra^B
\eeax
with $a_0^2={a'}_0^2\geq\frac12$, $a_1'a_2\neq a_1a_2'$, $\sum_i a_i^2=\sum_i {a'}_i^2=1$, $a_0>a_1\geq a_2$, $a'_0>a'_1\geq a'_2$, 
and
\bea\label{eg3}
|\Phi\ra=\frac{1}{\sqrt{2}}\left(|\psi_0\ra^{AB}|0\ra^C+|\psi_1\ra^{AB}|1\ra^C\right). 
\eea
After tracing over subsystems we are left with
\beax
\rho^{AB}&=&\frac12\left(|\psi_0\ra\la\psi_0|^{AB}+|\psi_1\ra\la\psi_1|^{AB}\right),\\
\rho^{AC}&=&\frac12\left[\left( a_0^2|0\ra\la0|^A+a_1^2|1\ra\la1|^A+a_2^2|2\ra\la2|^A\right)\ot|0\ra\la0|^C\right. \\
&&+\left( {a'}_0^2|0\ra\la0|^A+{a'}_1^2|1\ra\la1|^A+{a'}_2^2|2\ra\la2|^A\right)\ot|1\ra\la1|^C \\
&&+ \left( a_1a'_2|1\ra\la2|^A+a_2a'_1|2\ra\la1|^A\right)\ot|0\ra\la1|^C\\
&&+ \left. \left( a_1a'_2|2\ra\la1|^A+a_2a'_1|1\ra\la2|^A\right)\ot|1\ra\la0|^C
\right],\\
\rho_0^A&=&a_0^2|0\ra\la0|^A+a_1^2|1\ra\la1|^A+a_2^2|2\ra\la2|^A,\\
\rho_1^A&=&a_0^2|0\ra\la0|^A+{a'}_1^2|1\ra\la1|^A+{a'}_2^2|2\ra\la2|^A
\eeax
and
\beax
\rho^A&=&a_0^2|0\ra\la0|^A+\frac12({a}_1^2+{a'}_1^2)|1\ra\la1|^A
+\frac12({a}_2^2+{a'}_2^2)|2\ra\la2|^A,
\eeax
where $\rho_{0,1}^A=\tr_b|\psi_{0,1}\ra\la\psi_{0,1}|^{AB}$.
From here it follows that $E_2(|\Phi\ra^{A|BC})=1-a_0^2$.
We next show that $E_2(\rho^{AB})=E_2(|\Phi\ra^{A|BC})$ but $E_2(\rho^{AC})>0$, namely,
$E_2$ is not monogamous.
For any pure state ensemble of $\rho^{AB}=\sum_ip_i|\phi_i\ra\la\phi_i|^{AB}$, 
we have
\beax
p_i|\phi_i\ra^{AB}=\frac{1}{\sqrt{2}}\left( u_{i0}|\psi_0\ra^{AB}+u_{i1}|\psi_1\ra^{AB}\right) 
\eeax
for any $i$, where $|u_{i0}|^2+|u_{i1}|^2\leq 1$, which yields
the largest eigenvalue of $\sigma_i^A=\tr_B|\phi_i\ra\la\phi_i|^{AB}$
is always $a_0$. Thus 
\beax 
E_2(\rho^{AB})=E_2(|\Phi\ra^{A|BC})=1-a_0^2
\eeax as desired.
On the other hand,
we let $|x\ra^{AC}=a_1|1\ra^A|0\ra^C+a_2'|2\ra^A|1\ra^C$
and $|y\ra^{AC}=a_2|2\ra^A|0\ra^C+a_1'|1\ra^A|1\ra^C$,
then
\beax
\rho^{AC}&=&a_0^2|0\ra\la0|^A\ot(|0\ra\la0|^C+|1\ra\la1|^C)
+\frac12|x\ra\la x|^{AC}+\frac12|y\ra\la y|^{AC}.
\eeax
It is easy to see that $\rho_{AC}^{T_A}$ is not positive whenever $a_1'a_2\neq a_1a_2'$, 
and thus $E_2(\rho^{AC})>0$, here $T_X$ 
denotes the partial transpose transformation with respect to
the subsystem $X$.

If the reduced subsystem is two-dimensional, we consider the three-qubit case with no loss of generality.
Any pure state $|\psi\rangle$ in $\C^{2}\otimes \C^{2}\otimes \C^{2} $ can be expressed as~\cite{Acin2000prl}
\beax
|\psi\rangle^{ABC} &=&\lambda_{0}|000\rangle+\lambda_{1}e^{{\rm i}\varphi}|100\rangle+\lambda_{2}|101\rangle
+\lambda_{3}| 110\rangle+\lambda_{4}|111\rangle
\eeax
up to local unitary transformation,
where $\lambda_{i}\geq0$, $0 \leq \varphi \leq \pi$, $\sum_{i}{\lambda_{i}}^{2}=1$.
The reduced states 
$\rho^{AB}=p|x_1\ra\la x_1|+(1-p)|x_2\ra\la x_2|$ with
$\sqrt{p}|x_1\ra=\lambda_2|10\ra+\lambda_4|11\ra$ and $\sqrt{(1-p)}|x_2\ra=\lambda_0|00\ra+\lambda_1e^{{\rm i}\varphi}|10\ra+\lambda_3|11\ra$,
and
\beax
\rho^{A}=
\left(
\begin{array}{cc}
	\lambda_{0}^2 & \lambda_0\lambda_1e^{-{\rm i}\varphi}\\
	\lambda_0\lambda_1e^{{\rm i}\varphi}& \lambda_1^2+\lambda_2^2+\lambda_3^2+\lambda_4^2
\end{array}
\right).
\eeax
It is straightforward that
(1) $|\psi\ra$ is genuinely entangled if and only if
$\lambda_0> 0$, $\lambda_2^2+\lambda_4^2>0$ and $\lambda_3^2+\lambda_4^2>0$,
(2) $\rho^{AB}$ is separable iff $\lambda_3=0$,
and (3) $\rho^{AC}$ is separable iff $\lambda_2=0$.
If $E_2(|\psi\ra^{A|BC})=E_2(\rho^{AB})$, then
\beax 
E_2(\rho^{AB})=\sum_kp_kE_2(|\phi_k\ra)
\eeax 
for any
$\rho^{AB}=\sum_kp_k|\phi_k\ra\la\phi_k|$ according to Corollary 5 in Ref.~\cite{GG}.
This leads to
the minimal eigenvalue of 
\beax 
(1-p)\tr_B|x_2\ra\la x_2|=\left(
\begin{array}{cc}
	\lambda_{0}^2 & \lambda_0\lambda_1e^{-{\rm i}\varphi}\\
	\lambda_0\lambda_1e^{{\rm i}\varphi}& \lambda_1^2+\lambda_3^2
\end{array}
\right)
\eeax
coincides with that of $\rho^A$, which yields either
$\lambda_2=\lambda_4=0$, or $\lambda_1=0$ and $\lambda_0\leq \lambda_3$. That is, $\rho^{AC}$ could be entangled. Therefore
$E_2$ is still not monogamous whenever the reduced subsystem is two dimensional.

Let $\lambda_{\min}$ be the minimal positive Schmidt coefficient of $|\psi\ra$.
We define
\bea
E_{\min}(|\psi\ra)=\left\lbrace \begin{array}{ll}~\lambda_{\min}^2,& \lambda_{\min}<1,\\
	~0, &\lambda_{\min}=1
\end{array}\right. 
\eea 
for pure state and then define by means of the convex-roof extension for mixed state.
Denoting by 
\bea
\|\rho\|_{\min}=\left\lbrace \begin{array}{ll}~\lambda_{\min}^2,& \lambda_{\min}<1,\\
	~0, &\lambda_{\min}=1.
\end{array}\right. 
\eea  
it turns out that
\beax
E_{\min}(|\psi\ra)=h(\rho^A)=\|\rho^A\|_{\min}.
\eeax
We call $E_{\min}$ the \textit{minimal partial-norm of entanglement}, which reflects as the minimal case of the partial-norm. 
It is clear that $E_{\min}(\rho)=0$ iff $\rho$ is separable. Let $\delta(\rho)=(\delta_1, \delta_2, \dots, \delta_d)$ for any state $\rho\in\mS$ with $\dim\mH=d$, where $\delta_i$s are the eigenvalues,
$\delta_1\geq\delta_2\geq\dots\geq\delta_d$.
The concavity of $h$ is clear since
\beax 
\delta[t\rho+(1-t)\sigma]\prec t\delta(\rho)+(1-t)\delta(\sigma),
\eeax
which implies 
\beax
\|t\rho+(1-t)\sigma\|_{\min}\geq t\|\rho\|_{\min}+(1-t)\|\sigma\|_{\min},
\eeax
where ``$\prec$'' is the majorization relation between probability distributions.
Thus $E_{\min}$ is an entanglement monotone.

By now, except for the convex-roof extension of the negativity $N$~\cite{Lee}, denoted by $N_F$, all the reduced functions of convex-roof extended entanglement monotones in previous literature are shown to be strictly concave. 
Here
$N$ is defined as~\cite{Vidal02} 
$N(\rho)=\sum_i\mu_i$ with $\mu_i$s are the eigenvalues of the negative part of $\rho^{T_A}$. 
In order to show that the reduced function of $N_F$, denoted by $h_N$, is strictly concave. We give the following statement at first, which is a complementary of Vidal~\cite[Theorem 2]{Vidal2000}.

\begin{pro}{\label{pro1}}
	Let $E$ be an entanglement measure with the reduced function $h$ defined as Eq.~(\ref{h}). 
	If $E$ is an entanglement monotone, then $h$ is concave.	
\end{pro}

\begin{proof}
Let $\rho$ and $\sigma$ be any given two states in $\mS^A$, $0\leq t\leq1$.
	Taking $|\psi\ra^{AB}$ and $|\phi\ra^{AB}$ in $\mH^{AB}$ such that
	$\rho=\tr_B|\psi\ra\la\psi|^{AB}$ and $\sigma=\tr_B|\phi\ra\la\phi|^{AB}$,
	we let
	\beax
	|\Psi\ra^{ABC}=\sqrt{t}|\psi\ra^{AB}|0\ra^{C}+\sqrt{1-t}|\phi\ra^{AB}|1\ra^{C}
	\eeax
	be a pure state in $\mH^{ABC}$.
	Consider a LOCC $\{I^A\ot I^B\ot |0\ra\la0|^C, I^A\ot I^B\ot |1\ra\la1|^C\}$ acting on $|\Psi\ra^{ABC}$, we obtain
	the output 
	\beax
	\left\lbrace t|\psi\ra\la\psi|^{AB}\ot|0\ra\la0|^C, (1-t)|\phi\ra\la\phi|^{AB}\ot|1\ra\la1|^C\right\rbrace,
	\eeax 
	where $I^{A,B}$ is the identity operator acting on $\mH^{A,B}$.
	This leads to
	\beax
	E(|\Psi\ra^{A|BC})\geq tE(|\psi\ra^{AB}|0\ra^C)+(1-t)E(|\phi\ra^{AB}|1\ra^C)
	\eeax
	since $E$ is an entanglement monotone, which is equivalent to
	\beax
	h(t\rho+(1-t)\sigma)\geq th(\rho)+(1-t)h(\sigma),
	\eeax	
	that is, $h$ is concave. 		
\end{proof}

By Proposition~\ref{pro1}, $N_F$ is an entanglement monotone since $N$ is an entanglement monotone and thus the reduced function $h_N$ is concave. Note here that, in Ref.~\cite{Lee}, there is a gap in the proof of the concavity of $h_N$: the second inequality of the last part in page 2 is wrong since 
${|\phi_k\rangle}$ is not necessarily a basis (i.e., it is just an orthogonal set but not complete) in general.
We show that $h_N$ is strictly concave as well. 
We assume to obtain a contradiction that $h_N$ is not strictly concave.
Then there exists $\rho^A=p\rho_1^A+(1-p)\rho_2^A\in\mS^A$ with ${\rm spec}(\rho_1^A)\neq{\rm spec}(\rho_2^A)$, but
$h_N(\rho^A)=ph_N(\rho_1^A)+(1-p)h_N(\rho_2^A)$, here ${\rm spec}(X)$ denotes the spectrum of $X$.
Let 
\beax
\rho^{AB}=p|\psi_1\ra\la\psi_1|^{AB}+(1-p)|\psi_2\ra\la\psi_2|^{AB}
\eeax
with $|\psi_i\ra^{AB}=\sum_j\lambda_{ij}|e_{ij}\ra^A|e_{ij}\ra^B$ is the Schmidt decomposition of $|\psi_i\ra^{AB}$, $i=1$, 2,
where $\tr_B|\psi_i\ra\la\psi_i|^{AB}=\rho_i^A$, and
\beax
\la e_{ij}|e_{kl}\ra^{B}=\delta_{ik}\delta_{jl}.
\eeax
We take 
\beax
|\widetilde{\Psi}\ra^{ABC}=\sqrt{p}|\psi_1\ra^{AB}|0\ra^C+\sqrt{1-p}|\psi_2\ra^{AB}|1\ra^C,
\eeax 
then for any ensemble of $\rho^{AB}=\sum_kq_k|\phi_k\ra\la\phi_k|^{AB}$,
\beax
&&\sum_kq_kN(|\phi_k\ra^{AB})
=\sum_kq_kh_N(\rho_k^A)\\
&\geq&\sum_k\left[ p|u_{k1}|^2h_N(\rho_1^A)+(1-p)|u_{k2}|^2h_N(\rho_2^A)\right] \\
&=&ph_N(\rho_1^A)+(1-p)h_N(\rho_2^A),
\eeax
where 
\beax
\sqrt{q_k}|\phi_k\ra^{AB}=u_{k1}\sqrt{p}|\psi_1\ra^{AB}+u_{k2}\sqrt{1-p}|\psi_2\ra^{AB}.
\eeax
It turns out that
$N(\rho^{AB})=N(|\widetilde{\Psi}\ra^{A|BC})$.
But $|\widetilde{\Psi}\ra^{ABC}$ does not admit the form $|\psi\ra^{AB_1}|\psi\ra^{B_2C}$ up to some local unitary operation, 
where $B_1B_2$ means $\mH^B$ has a subspace isomorphic to $\mH^{B_1}\otimes\mH^{B_2}$ and up to local unitary on system $B_1B_2$, which contradicts with Theorem 3 in~\cite{Hehuan}.
Thus $h_N$ is strictly concave.
That is, all the reduced functions of the monogamous entanglement monotones so far are strictly concave.

We now go back to discuss the monogamy of $E_{\min}$. Clearly, if the reduced system is two-dimensional,
then $E_{\min}=E_2$, which is not monogamous.
For higher dimensional case, 
we consider a pure state as in Eq.~(\ref{eg3}) just by replacing $a_0^2={a'}^2_0\geq\frac12$, $a_0>a_1\geq a_2$, $a'_0>a'_1\geq a'_2$, 
with $a_0=a'_0$, $a_1\geq a_2>a_0$, $a'_1\geq a'_2>a'_0$,
from which one can conclude that $E_{\min}$ is not monogamous.

However, $E_{\min}$ does not achieve the maximal value for the maximally entangled state.
For making up the disadvantages, we can define
\bea
E'_{\min}(|\psi\ra)=\left\lbrace \begin{array}{ll}~\lambda_{\min}^2S_r(|\psi\ra),& \lambda_{\min}<1,\\
	~0, &\lambda_{\min}=1,
\end{array}\right. 
\eea 
for pure state and then define by means of the convex-roof extension for mixed state,
where $S_r(|\psi\ra)$ denotes the Schmidt rank of $|\psi\ra$. We call it the \textit{reinforced minimal partial-norm of entanglement}.
$E'_{\min}$ is equal to $2E_{\min}$
for any $2\ot n$ state.
In such a case, $E'_{\min}$ reaches the maximal quantity for the maximally entangled state but not only for these states.
In addition, it is easy to follow that $E'_{\min}$ is also an entanglement monotone and is not monogamous.

Let $|\psi\ra$, $|\phi\ra$, $|\varphi\ra$, $|\xi\ra$, and $|\zeta\ra$ be 
pure states with the reduced states, respectively, are
$\diag(2/3, 1/6, 1/6)$, $\diag(1/3, 1/3, 1/3)$,
$\diag(3/5, 2/5, 0)$, $\diag(2/5, 2/5, 15)$,
and $\diag(4/5, 1/5, 0)$.
Then  we arrive at
\beax 
E_2(|\varphi\ra)<E_2(|\phi\ra)
~~\mbox{but} ~~
E_{\min}(|\phi\ra)<E_{\min}(|\varphi\ra),
\eeax
\beax
E_2(|\varphi\ra)<E_2(|\xi\ra)
~~\mbox{but} ~~ E'_{\min}(|\xi\ra)<E'_{\min}(|\varphi\ra),
\eeax
\beax E_{\min}(|\psi\ra)<E_{\min}(|\zeta\ra)
~~\mbox{but} ~~ E'_{\min}(|\zeta\ra)<E'_{\min}(|\psi\ra).
\eeax
That is, these three measures are not equivalent to each other. 

The maximal value of $E_2$ is $(d-1)/d$.
We thus, in order to get a normalized measure, replace $E_2$ by $dE_2/(d-1)$. Hereafter the notation $E_2$ refers to the normalized one.
For the $2\ot n$ system, $E_2$ coincides with $E'_{\min}$ but not for $m\ot n$ system with $2<m\leq n$.
For any pure state $|\psi\ra$ with Schmidt numbers $p$ and $1-p$ in $2\ot n$ system, $p\leq1/2$, it is immediate that
\beax
E_2(|\psi\ra)=2E_{\min}(|\psi\ra)=E'_{\min}(|\psi\ra)=2p,
\eeax and 
\beax
\tau(|\psi\ra)=2p(1-p).
\eeax Here, $\tau$ is the tangle, which is defined as the square of concurrence, i.e.,
$\tau(|\psi\ra)=2(1-\tr\rho_A^2)$, $\rho_A=\tr_B|\psi\ra\la\psi|$.
That is $E_2\geq\tau$ and both of them are monotonically increasing with $0\leq p\leq1/2$.

\begin{figure}
	\includegraphics{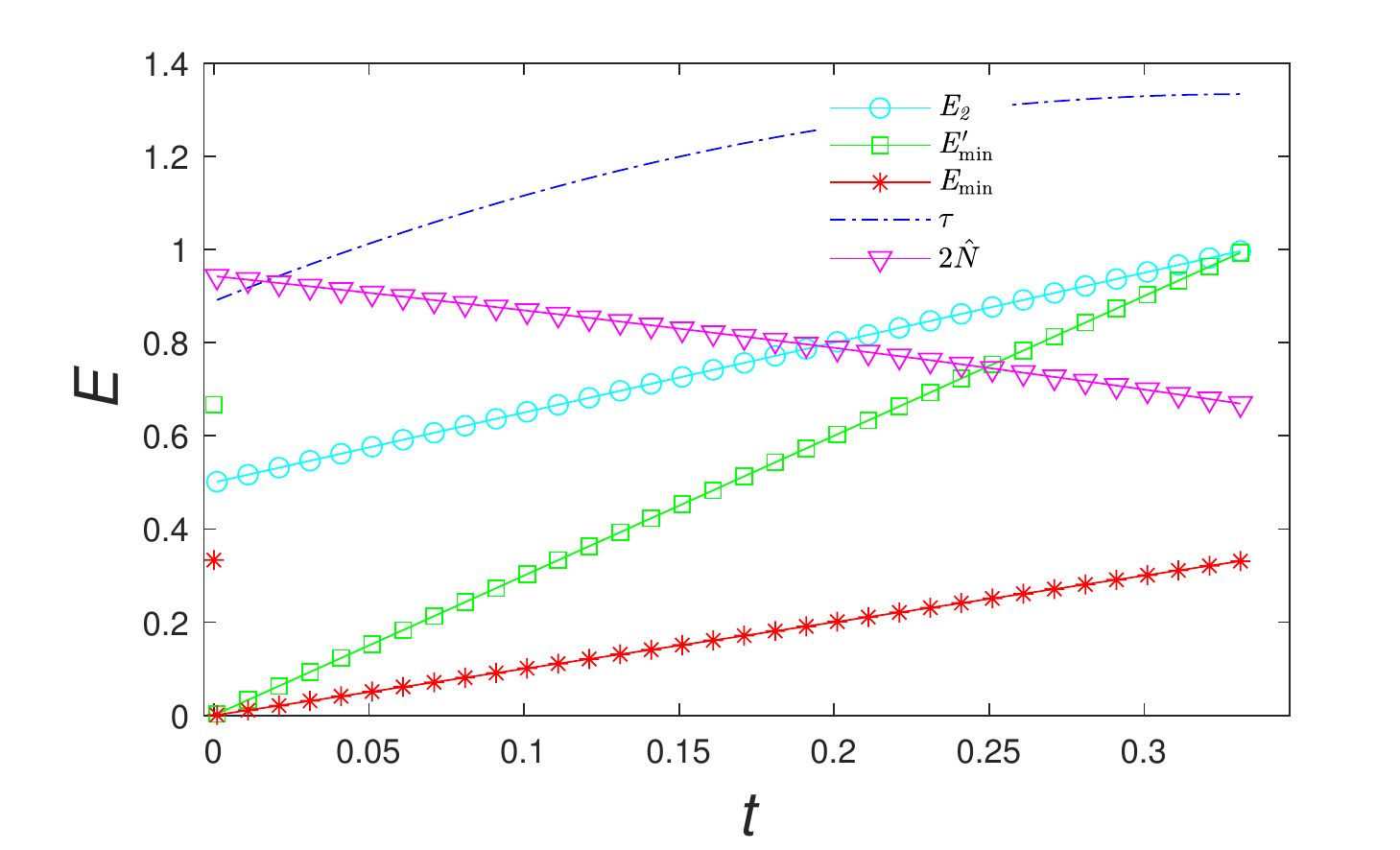}
	\caption{\label{fig1}(color online). Comparing $E_2$, $E'_{\min}$ with the tangle $\tau$ for $|\psi\ra$ with Schmidt numbers $(\sqrt{2/3-t}, \sqrt{1/3}, \sqrt{t})$, $0\leq t\leq1/3$.}
\end{figure}

\begin{figure}
	\includegraphics{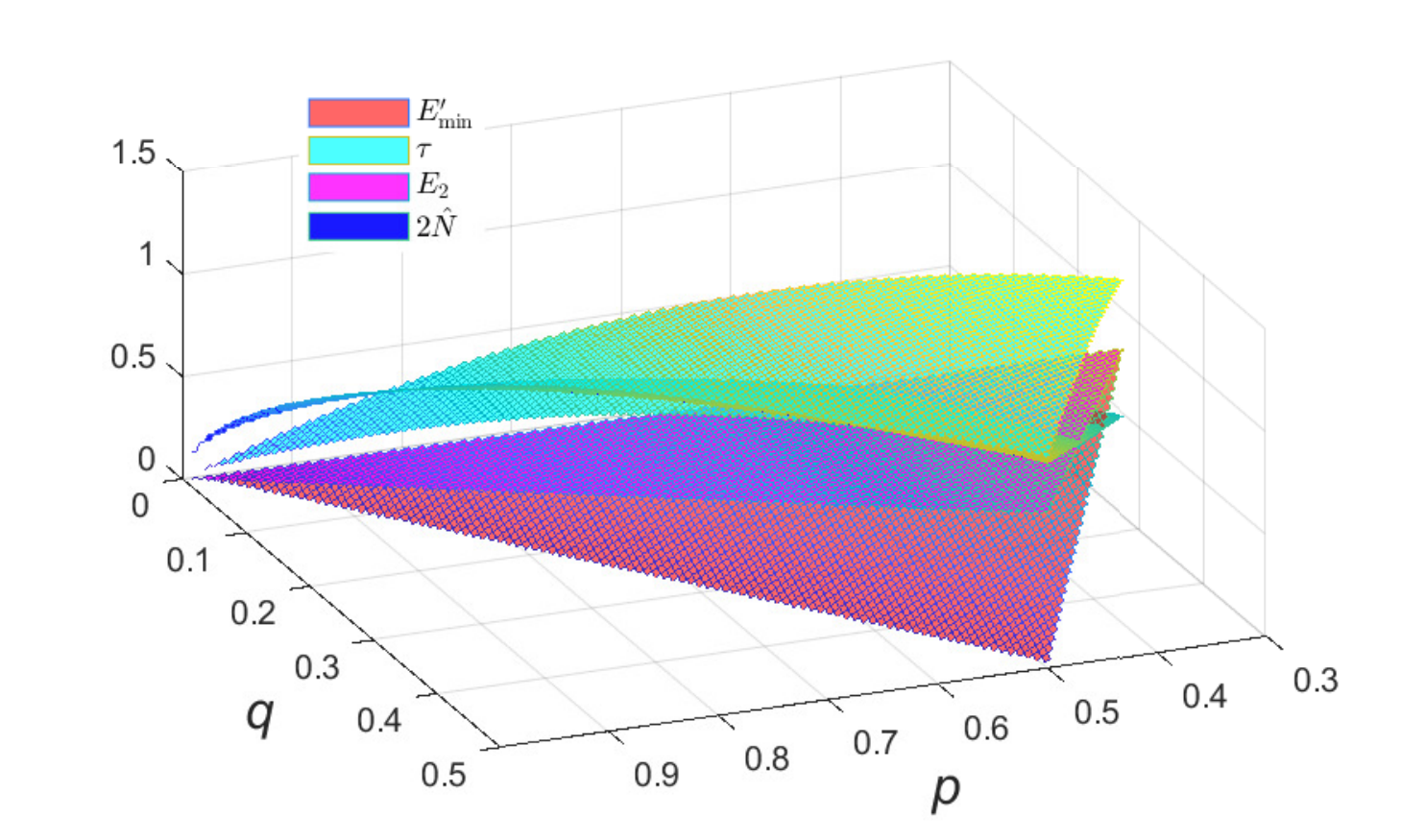}
	\caption{\label{fig2}(color online). Comparing $E_2$, $E'_{\min}$ with the tangle $\tau$ for $|\phi\ra$ with Schmidt numbers $(\sqrt{p}, \sqrt{q}, \sqrt{1-p-q})$, $p+q<1$. }
\end{figure}

We now compute these three entanglement monotones for the qutrit-qutrit pure state and then compare them with tangle.
It can be easily calculated since they are homogeneous. We consider $|\psi\ra\in\C^3\ot\C^3$ with Schmidt numbers $(\sqrt{2/3-t}, \sqrt{1/3}, \sqrt{t})$, $0\leq t\leq1/3$,
and $|\phi\ra\in\C^3\ot\C^3$ with Schmidt numbers $(\sqrt{p}, \sqrt{q}, \sqrt{1-p-q})$ for illustration purposes, $p\geq q$.
The behaviours of theses quantities for these two states are depicted in Fig.~\ref{fig1} and Fig.~\ref{fig2}, respectively. In the case of $t=0$, $E'_{\min}(|\psi\ra)=2/3=2E_{\min}(|\psi\ra)$ and $\tau(|\psi\ra)=8/9$. 
For the case of $p+q=1$, $E_2(|\psi\ra)=3q/2<E'_{\min}(|\psi\ra)=2q$. That is, $E_{\min}$ and $E'_{\min}$ are not continuous and are not equivalent.

The upper bounds of these quantities can be easily derived.
Let $\rho$ be a state in $\mS^{AB}$, and $E_2(\rho)=\sum_ip_iE_2(|\psi_i\ra)$,
then
\beax &&E_2(\rho)=\sum_ip_iE_2(|\psi_i\ra)\\
&=&\frac{d}{d-1}\sum_ip_i\left(1- \|\rho_i^A\|\right) \\
&=&\frac{d}{d-1}\left[ 1-\left(\sum_ip_i\|\rho_i^A\|\right)\right] \\
&\leq& \frac{d}{d-1}\left( 1-\|\sum_ip_i\rho_i^A\|\right) \\
&=&\frac{d}{d-1}\left( 1-\|\rho^A\|\right).
\eeax 
That is
\bea
E_2(\rho)\leq\frac{d}{d-1}\min\{1-\|\rho^A\|, 1-\|\rho^B\|\}.
\eea
Analogously,
\bea
E_{\min}(\rho)\leq \min\{\|\rho^A\|_{\min}, \|\rho^B\|_{\min}\}
\eea
and 
\bea
E'_{\min}(\rho)\leq \min\{r_A\|\rho^A\|_{\min}, r_B\|\rho^B\|_{\min}\},
\eea
where $r_{A,B}$ is the rank of $\rho^{A,B}$.

When $k\geq3$, $E_k$ is not a faithful entanglement monotone, and it is not monogamous either. 
Another entanglement measure that lack of investigating the monogamy is the Schmidt number, which is regarded as a universal entanglement measure~\cite{Sperling2011}, defined by~\cite{Terhal1999}
\bea
S_r(\rho)=\min_{p_i,|\psi_i\ra}\max_{|\psi_i\ra}S_r(|\psi_i\ra),
\eea
where the minimum is taken over all decomposition $\rho=\sum_ip_i|\psi_i\ra\la\psi_i|$.
It is also not monogamous since both the Schmidt number of $|W\ra=\frac{1}{\sqrt{3}}(|100\ra+|010\ra+|001\ra)$ and that of its two reduced states are 2.

In addition, let $\rho^{T_A}$ be the partial transpose of $\rho$, one may consider the partial-norm of the negative part of $\rho^{T_A}$, $N\rho^-$. For example, 
we take
\bea
\hat{N}(\rho)=\|N\rho^-\|.
\eea
We call it \textit{partial negativity} hereafter.
Take $\rho=|\psi\ra\la\psi|$ with
$|\psi\ra=\sum_j\lambda_j|e_j\ra^A|e_j\ra^B$ as the 
Schmidt decomposition of $|\psi\ra$.
Then 
$\hat{N}(|\psi\ra)=\lambda_1\lambda_2$,
and the corresponding reduced function is 
\bea
\hat{h}(\rho^A)=\sqrt{\delta_1\delta_2},
\eea
where $\delta_1=\lambda_1^2$, $\delta_2=\lambda_2^2$.
$\hat{N}$ can still be regarded as a kind of partial norm as $\sqrt{\delta_1\delta_2}\leq\delta_1=\|\rho^A\|$, in other words, $\hat{N}$ is also a kind of partial norm of entanglement.
By definition, $\hat{N}(|\psi\ra^{ab})=0$ if and only if it is separable, and
\beax
0<\hat{N}(\rho)\leq N(\rho)
\eeax
for any non-positive partial transpose state $\rho$.
A simple comparison between $\hat{N}$ and $E_2$, $E_{\min}$, $E'_{\min}$ are given in Fig.~\ref{fig1} and Fig.~\ref{fig2}, which indicate that they are not equivalent to each other.
For the two-qubit case, $2\hat{N}_F$ coincides with the $G$-concurrence~\cite{Gour2005pra}.
We conjecture that $\hat{h}$ is concave~\cite{Suppl}.
$\hat{h}$ is strictly concave on $\mS(\mH)$ with $\dim\mH=2$ since it reduced to an elementary symmetric function~\cite[p.~116]{Marshall}, but it is not true for the higher dimensional case.
In order to see this, we take 
\beax
\rho=\left( \begin{array}{ccc}
	1/3&0&0\\
	0&1/3&0\\
	0&0&1/3
\end{array}\right),\quad
\sigma=\left( \begin{array}{ccc}
	1/2&0&0\\
	0&1/2&0\\
	0&0&0
\end{array}\right),
\eeax
which yields $\hat{h}(\frac12\rho+\frac12\sigma)=\frac12\hat{h}(\rho)+\frac12\hat{h}(\sigma)$.
We now assume that $\hat{N}$ is an entanglement monotone, then we can conclude the following.

\begin{theorem}
	$\hat{N}$ and $\hat{N}_F$ are not monogamous whenever the reduced subsystem has dimension greater than 2.
\end{theorem}

We show this statement by a counter-example. Let
\bea\label{eg2}
|\widetilde{\Omega}\ra^{ABC}=\lambda_0|0\ra^A|00\ra^{BC}+\lambda_1|1\ra^A|10\ra^{BC}
+\lambda_2|2\ra^A|11\ra^{BC}
\eea
with $\lambda_0\geq \lambda_1\geq \lambda_2>0$,
it turns out that
\beax
\hat{N}(|\widetilde{\Omega}\ra^{A|BC})=\hat{N}(\rho^{AB})=\lambda_0\lambda_1
\eeax 
but
\beax
\hat{N}(\rho^{AC})=\lambda_1\lambda_2>0.
\eeax
That is, $\hat{N}$ is not monogamous.
Moreover, from this example, we can also get $\hat{N}_F$ is not monogamous either in light of $\hat{N}\leq\hat{N}_F$.

Analogous to that of the logarithmic negativity, we define the \textit{logarithmic partial negativity} by
\bea
\hat{N}_l(\rho)=\log_2[\hat{N}(\rho)+1].
\eea
It is straightforward that $\hat{N}_l$ is not convex.
For any LOCC acting on $\rho^{ab}$ that leaves the output states $\{p_i\sigma_i\}$, 
we have
\beax
\sum_ip_i\hat{N}_l(\sigma_i)=\sum_ip_i\log_2x_i\leq\log_2\sum_ip_ix_i
\leq\log_2[\hat{N}(\rho)+1]=\hat{N}_l(\rho)
\eeax
since $\log_2$ is concave and $\hat{N}$ is non-increasing on average under LOCC by assumption,
where $x_i=\hat{N}(\sigma_i)+1$.
Therefore it is also an entanglement monotone and is not monogamous (hereafter, we still call it an entanglement monotone even though it is not convex as in Ref.~\cite{Plenio2005prl}).

In sum, for the sake of distinguishing these entanglement monotones so far in the sense of monogamy law, we suggest the term \textit{informationally complete entanglement monotone}, which means that its reduced function is related to all its eigenvalues. For example, the entanglement of formation is informationally complete since the von Neumann entropy is defined on all of the eigenvalues which include all the information of the entanglement, but $E_2$, $E_{\min}$, $E'_{\min}$, $\hat{N}$, $\hat{N}_F$, and $\hat{N}_l$ are not the case except for the two-dimensional case since they just capture ``partial information'' of the entanglement.
The worst one is the Schmidt number, which reflects the least information of the entanglement, and of course is not informationally complete.
Our discussion supports that, 
for an entanglement monotone $E_F$ with reduced function $h$,
$E_F$ is monogamous if and only if it is informationally complete, and in turn, 
iff $h$ is strictly concave (the ``if'' part is proved~\cite{GG2019}).
So the axiomatic definition of an entanglement monotone should be improved as follows. 
Let $E$ be a nonnegative function on $\mS^{AB}$ with $E(|\psi\ra)=h(\rho^A)$ for pure state. We call $E$ a \textit{strict entanglement monotone} if {(i)} $E(\sigma^{AB})=0$ for any 
separable density matrix $\sigma^{AB}\in\mS^{AB}$, {(ii)} 
$E$ behaves monotonically decreasing under LOCC on average, and {(iii)} the reduced function $h$ is strictly concave.
We use henceforth the term strict entanglement monotone to distinguish it from the previous entanglement monotone.

With such a spirit,  except for $E_2$,  $E_{\rm min}$, $E'_{\rm min}$, $\hat{N}$, $\hat{N}_F$, $\hat{N}_l$ and the Schmidt number, all the previous entanglement monotones that are shown to be monogamous or monogamous on pure states are strict entanglement monotones, these include the original entanglement of formation, negativity, the squashed entanglement~\cite{Christandl2004jmp}, the convex-roof extension of negativity, tangle, concurrence, the relative entropy of entanglement~\cite{Vedral1998pra}, $G$-concurrence, the Tsallis entropy of entanglement, the conditional entanglement of mutual information~\cite{Yang2008prl}, and the entanglement measures induced by the fidelity distances, etc. 
However, it still remains unknown that whether or not the non convex-roof extended strict entanglement monotones in literature are monogamous in addition to the squashed entanglement. We conjecture that all the informationally complete entanglement monotones are monogamous.

As a by-product, we can obtain new coherence measures from the reduced function $h$ of $E_2$, $E_{\rm min}$ and $E'_{\rm min}$, respectively. Let
\bea
C_h(|\psi\ra)=h(x_0, x_1, \dots, x_{d-1})
\eea
for pure state $|\psi\ra=\sum_ix_i|i\ra$ under the reference basis $\{|i\ra\}_{i=0}^{d-1}$,
and by the convex-roof extension for mixed state, i.e.,
\beax
C_h(\rho)=\min_{p_j,|\psi_j\ra}\sum_jp_jC_h(|\psi_j\ra),
\eeax
where the minimum is taken over all decomposition $\rho=\sum_jp_j|\psi_j\ra\la\psi_j|$. 
It turns out that (i) $h(1, 0, \cdots, 0)=0$, (ii) $h(\pi(x_0, x_1, \dots, x_{d-1}))=h(x_0, x_1, \dots, x_{d-1})$
for any permutation $\pi$ and any $(x_0, x_1, \dots, x_{d-1})$, and (iii) $h$ is concave.
This reveals that $C_h$ is a well-defined coherence measure according to Theorem 1 in Ref.~\cite{Du2015qic}. 
Also notice here that, the associated function $h$ of all the previous coherence measures defined by means of the convex-roof extension are strictly concave, which are different from $C_h$.


\ack{
This work is supported by the National Natural Science Foundation of
China under Grant No.~11971277 and the Fund Program for the Scientific Activities of Selected
Returned Overseas Professionals in Shanxi Province under Grant No.~20220031.
}	



\section*{References}

\begin{thebibliography} {99}
	

\bibitem{Nielsen} Nielsen M A, Chuang I L 2000 \textit{Quantum Computatation and
	Quantum Information} (Cambridge: Cambridge University Press)

\bibitem{Schrodinger1935} Schr\"{o}dinger E 1935 Discussion of Probability Relations between Separated Systems
\textit{Proc. Cambridge Philos. Soc.} \textbf{31} 555

\bibitem{Schrodinger1936} Schr\"{o}dinger E 1936
Probability relations between separated systems
\textit{Proc. Cambridge Philos. Soc.} \textbf{32} 446


\bibitem{Bennett1993teleporting} Bennett C H, Brassard G, Cr\'{e}peau C, Jozsa R,
Peres A and Wootters W K 1993 
Teleporting an unknown quantum state via dual classical and Einstein-Podolsky-Rosen channels
\textit{Phys. Rev. Lett.} \textbf{70} 1895

\bibitem{Zhang2006experimental} Zhang Q, Goebel A, Wagenknecht C, Chen Y A,
Zhao B, Yang T, Mair A, Schmiedmayer J and Pan J W 2006 
Experimental quantum teleportation of a two-qubit composite system,
\textit{Nat. Phys.} \textbf{2} 678

\bibitem{Bennett1992communication} Bennett C H and Wiesner S J 1992
Communication via one-and two-particle operators on Einstein-Podolsky-Rosen states
\textit{Phys. Rev. Lett.} \textbf{69} 2881

\bibitem{Bennett1996prl} Bennett C H, Brassard G, Popescu S, Schumacher B,  Smolin J A and Wootters W K 1996
Purification of noisy entanglement and faithful teleportation via noisy channels,
\textit{Phys. Rev. Lett.} \textbf{76} 722

\bibitem{Terhal2004} Terhal B 2004
Is entanglement monogamous?
\textit{IBM J. Res. Dev.} \textbf{48} 71

\bibitem{Osborne} Osborne T J and Verstraete F 2006 
General monogamy inequality for bipartite qubit entanglement
\textit{Phys. Rev. Lett.} \textbf{96} (22) 220503  

	\bibitem{Horodecki2009} Horodecki R, Horodecki P, Horodecki M and Horodecki K 2009
Quantum entanglement
\textit{Rev. Mod. Phys.} \textbf{81} 865 

\bibitem{Bennett2014} Bennett C H 2014
in \textit{Proceedings of the FQXi 4th International Conference, Vieques Island, Puerto Rico}, 
http://fqxi.org/conference/talks/2014.

\bibitem{Toner} Toner B 2009 
Monogamy of non-local quantum correlations
\textit{Proc. R. Soc. A} \textbf{465} 59

\bibitem{Seevinck} Seevinck M P 2010  
Monogamy of correlations versus monogamy of entanglement
\textit{Quantum Inf. Process.} \textbf{9} 273

\bibitem{streltsov2012are} Streltsov A, Adesso G, Piani M \textit{et al} 2012
Are general quantum correlations monogamous?
\textit{Phys. Rev. Lett.} \textbf{109} 050503

\bibitem{Augusiak2014pra} Augusiak R, Demianowicz M, Paw\l{}owski M, Tura J and Ac\'{\i}n A 2014
Elemental and tight monogamy relations in nonsignaling theories,
\textit{Phys. Rev. A} \textbf{90} 052323

\bibitem{Ma2011} Ma X, Dakic B, Naylor W, Zeilinger A and Walther P 2011
Quantum simulation of the wavefunction to probe frustrated Heisenberg spin systems
\textit{Nat. Phys.} \textbf{7} 399

\bibitem{Brandao2013} Brandao F G S L and Harrow A W 2013 in \textit{Proceedings
	of the 45th Annual ACM Symposium on Theory of Computing}, http://dl.acm.org/citation.cfm?doid=2488608.2488718.

\bibitem{Garcia} Garc\'{\i}a-S\'{a}ez A and Latorre J I 2013 
Renormalization group contraction of tensor networks in three dimensions 
\textit{Phys. Rev. B} \textbf{87} 085130

\bibitem{Lloyd} Lloyd S and Preskill J 2014 
Unitarity of black hole evaporation in final-state projection models
\textit{J. High Energy Phys.} \textbf{08} 126

\bibitem{Pawlowski} Paw\l owski M 2010
Security proof for cryptographic protocols based only on the monogamy of Bell's inequality violations
\textit{Phys. Rev. A} \textbf{82} 032313

\bibitem{Vidal2000}  Vidal G 2000
Entanglement monotone
\textit{J. Mod. Opt.} \textbf{47} 355

\bibitem{Plenio2005prl} Plenio M B 2005
Logarithmic negativity: a full entanglement monotone that is not convex
\textit{Phys. Rev. Lett.} \textbf{95} 090503

\bibitem{Vedral1998pra} Vedral V and Plenio M B 1998
Entanglement measures and purification procedures
\textit{Phys. Rev. A} \textbf{57} 1619

\bibitem{Coffman} Coffman V, Kundu J and  Wootters W K 2000
Distributed entanglement
\textit{Phys. Rev. A} \textbf{61} 052306

\bibitem{Bai} Bai Y K, Xu Y F and Wang Z D 2014
General monogamy relation for the entanglement of formation in multiqubit systems
\textit{Phys. Rev. Lett.} \textbf{113} 100503 

\bibitem{Koashi} Koashi M and  Winter A 2004
Monogamy of quantum entanglement and other correlations
\textit{Phys. Rev. A} \textbf{69} 022309 

\bibitem{Luo2016pra} Luo Y, Tian T, Shao L H \textit{et al} 2016
General monogamy of Tsallis $q$-entropy entanglement in multiqubit systems
\textit{Phys. Rev. A} \textbf{93} 062340 

\bibitem{Dhar} Dhar H S, Pal A K, Rakshit D \textit{et al} 2017
Monogamy of quantum correlations-a review
\textit{In Lectures on General Quantum Correlations and their Applications}(Cham, Switzerland: Springer)

\bibitem{Hehuan} He H and Vidal G 2015
Disentangling theorem and monogamy for entanglement negativity
\textit{Phys. Rev. A} \textbf{91} 012339

\bibitem{GG} Gour G and  Guo Y 2018
Monogamy of entanglement without inequalities
\textit{Quantum} \textbf{2} 81

\bibitem{GG2019} Guo Y and Gour G 2019
Monogamy of the entanglement of formation
\textit{Phys. Rev. A} \textbf{99} 042305 

\bibitem{Bennett1996pra} Bennett C H, DiVincenzo D P, Smolin J A and Wootters W K 1996
Mixed-state entanglement and quantum error correction
\textit{Phys. Rev. A} \textbf{54} 3824

\bibitem{Rungta2003pra} Rungta P and Caves C M 2003
Concurrence-based entanglement measures for isotropic states,
\textit{Phys. Rev. A} \textbf{67} 012307 

\bibitem{Rungta} Rungta P, Bu\v{z}ek V, Caves C M, Hillery M and Milburn G J 2001
Universal state inversion and concurrence in arbitrary dimensions
\textit{Phys. Rev. A} \textbf{64} 042315

\bibitem{Gour2005pra} Gour G 2005 Family of concurrence monotones
and its applications \textit{Phys. Rev. A} \textbf{71} 012318

\bibitem{Kim2010pra} Kim J S 2010 
Tsallis entropy and entanglement constraints in multiqubit systems
\textit{Phys. Rev. A} \textbf{81} 062328

\bibitem{Guo2020qip} Guo Y, Zhang L and Yuan H 2020
Entanglement measures induced by fidelity-based distances
\textit{Quant. Inf. Process.} \textbf{19} 282 

\bibitem{Vidal1999pral} Vidal G 1999
Entanglement of pure states for a single copy
\textit{Phys. Rev. Lett.} \textbf{83} 1046

\bibitem{Acin2000prl} Ac\'{\i}n A \textit{et al}. 2000
Generalized Schmidt decomposition and classification of three-quantum-bit states
\textit{Phys. Rev. Lett.} \textbf{85} 1560

\bibitem{Vidal02} Vidal G and Werner R F 2002
Computable measure of entanglement
\textit{Phys. Rev. A} \textbf{65} 032314

\bibitem{Lee} Lee S, Chi D P, Oh S D and Kim J 2003
Convex-roof extended negativity as an entanglement measure for bipartite quantum systems
\textit{Phys. Rev. A} \textbf{68} 062304

\bibitem{Sperling2011} Sperling J and Vogel W 2011 
The Schmidt number as a universal entanglement measure
\textit{Phys. Scr.} \textbf{83} 045002 

\bibitem{Terhal1999} Terhal B M and Horodecki P 1999
Schmidt number for density matrices 
\textit{Phys. Rev. A} \textbf{61}
040301(R)

\bibitem{Suppl} We checked by many examples that it is true. Especially, if $[\rho, \sigma]=0$, $\hat{h}(\rho/2+\sigma/2)\geq\hat{\rho}/2+\hat{h}(\sigma)/2$ is always true. 

\bibitem{Marshall} Marshall A W, Olkin I and Arnold B C 2011
Inequalities: theory of majorization and its applications
\textit{New York: Academic press}

\bibitem{Christandl2004jmp} Christandl M and Winter A 2004
``Squashed entanglement'': an additive entanglement measure
\textit{J. Math. Phys. (N.Y.) }\textbf{45} 829

\bibitem{Yang2008prl} Yang D, Horodecki M and Wang Z D 2008
An additive and operational entanglement measure: conditional entanglement of mutual information
\textit{Phys. Rev. Lett.} \textbf{101} 140501

\bibitem{Du2015qic} Du S, Bai Z and Qi X 2015
Coherence measures and optimal conversion for coherent states
\textit{Quant. Inf. \& Comput.} \textbf{15} (15-16) 1307-1316 


	


	
	
\end{thebibliography}

\end{document}